\documentclass{aip-cp}
\usepackage[numbers]{natbib}
\usepackage{graphicx}

\newcommand{\sign}{{\rm sign}}

\begin{document}

\title{On the Keldysh Problem of Flutter Suppression
}

\author[aff1,aff2]{Gennady A. Leonov}
\eaddress{g.leonov@spbu.ru}
\author[aff1,aff3]{Nikolay V. Kuznetsov} 
\eaddress{Corresponding author: nkuznetsov239@gmail.com}

\affil[aff1]{
Saint-Petersburg State University, 7/9 Universitetskaya emb., Saint-Petersburg, 199034, Russia}
\affil[aff2]{Institute of Problems of Mechanical Engineering RAS,
61 Bolshoj pr. V.O., Saint-Petersburg, 199178, Russia}
\affil[aff3]{
University of Jyv\"{a}skyl\"{a},
P.O. Box 35  (Agora), Jyv\"{a}skyl\"{a}, FI-40014, Finland}

\maketitle

\begin{abstract}
This work is devoted to the Keldysh model of flutter suppression and rigorous approaches to its analysis.
To solve the stabilization problem in the Keldysh model we
use an analog of direct Lyapunov method for differential inclusions.
The results obtained here are compared
with the results of Keldysh obtained by
the method of harmonic balance (describing function method),
which is an approximate method for analyzing the existence of periodic solutions.
The limitations of the use of describing function method
for the study of systems with dry friction and stationary segment are demonstrated.

\end{abstract}

\section{INTRODUCTION}

The notion of flutter is traced back to the work of Lanchester \cite{Lanchester-1916},
where torsional vibrations of the tail of Handley Page 0/400 biplane bomber
were described. Flutter is a self-excited oscillation, often destructive,
wherein energy is absorbed from the airstream;
it is a complex phenomenon that must in general be completely eliminated
by design or prevented from occurring within the flight envelope \cite{GarrickR-1971}.
In the work by Parhomovsky and Popov \cite{ParhomovskyP-1971}  it is remarked that
about 150 crashes of new models of aircrafts caused by flutter
happened in German aviation during 1935-1943,
while in the Soviet Union the works of M.V.~Keldysh\footnote{
M.V.~Keldysh worked in TsAGI from 1931 till 1946;
in 1942 he got the State Stalin Prize ``for scientific works on the
prevention of aircraft destruction'';
from 1961 till 1975 he was the President of the Academy of Sciences of the USSR.
}
and his scientific school in TsAGI allowed to avoid the numerous accidents
that accompanied the development of aviation.
In this work\footnote{
See also the materials of our plenary lecture given at
the International Scientific Conference on Mechanics ``The Eighth Polyakhov's Reading'':
http://www.math.spbu.ru/user/nk/PDF/2018-PR-plenary-Flutter-suppression-Keldysh-model.pdf
}
we revisit the work by M.Keldysh on
the flutter suppression by dampers with a nonlinear characteristic,
published in 1944 \cite{Keldysh-1944}.

\section{\uppercase{Keldysh problem on the flutter suppression}} 
Follow \cite{Keldysh-1944}, first we consider the suppression of flutter
for a model with one degree of freedom
\begin{equation}\label{sysK2d}
  J\ddot x+kx = (-\lambda+h)\dot x-\varphi(\dot x),
  \quad \varphi(\dot x) = (\Phi+\kappa \dot x^2)\sign(\dot x),
\end{equation}
where $J>0$ --- the moment of inertia, $k>0$ --- stiffness,
$h\dot x$ --- an excitation force proportional to the angular velocity $\dot x$,
$f(\dot x)=\lambda\dot x+\varphi(\dot x)$ ---  nonlinear characteristic of hydraulic damper with dry friction,
$\Phi$ --- the dry friction coefficient,
$\lambda>0$ and $\kappa>0$ --- parameters of the hydraulic damper.
Following the mechanical sense, Keldysh defined $\sign(0)$
as a value from $[-\Phi,\Phi]$
and, thus, the discontinuous differential equation (\ref{sysK2d})
has a segment of equilibria (stationary segment).

Let
\(
  \mu=\lambda-h.
\)
Using the describing function method (DFM, harmonic balance),
which had been well developed by 1944  \cite{KrylovB-1937},
Keldysh formulated the following result:

{\it
\begin{enumerate}
\item If
\begin{equation}\label{2dKstab}
  -2.08 \sqrt{\Phi\kappa}< \mu ,
\end{equation}
then all trajectories of (\ref{sysK2d}) converge to the stationary segment;

\item If
\begin{equation}\label{2dK2lc}
  \mu < -2.08 \sqrt{\Phi\kappa},
\end{equation}
then there are two periodic trajectories (limit cycles)
$\approx a_{\pm}\cos(\omega t)$ with amplitudes
\[
 a_\pm(\mu) =\frac{3}{8\kappa}\sqrt{\frac{J}{k}}\bigg(\pi(\mu)\pm\sqrt{\pi^2(\mu)^2
 -\frac{32}{3}\kappa\Phi}\bigg).
\]
Other trajectories behave as follows.
The trajectories, emerging from infinity,
tend to the external limit cycle.
The domain between two limit cycles is filled with trajectories
unwinding from the internal (unstable) limit cycle and winding
onto external (stable) limit cycle.
The stability domain bounded by the internal limit cycle
is filled with trajectories tending to one of
the possible equilibrium on the stationary segment.
\end{enumerate}
}
\noindent It is well known that the classical DFM
is only an approximate method which gives the information on
the frequency and amplitude of periodic orbits and, in general,
may lead to wrong conclusions\footnote{
For example, well-known Aizerman's and Kalman's  conjectures
\cite{Aizerman-1949,Kalman-1957}
on the absolute stability of nonlinear control systems
are valid from the standpoint of DFM
which may explain why these conjectures were put forward.
Nowadays, there are known various counterexamples to these conjectures:
nonlinear systems, where the only equilibrium, which is stable,
coexists with a periodic oscillation, which is a hidden attractor
(see, e.g. surveys \cite{BraginVKL-2011,LeonovK-2013-IJBC} 
 and references within).
An attractor is called a self-excited
attractor if its basin of attraction intersects with
any open neighbourhood of an equilibrium;
otherwise it is called a hidden attractor \cite{LeonovK-2013-IJBC}.
Hidden attractors have been found in various engineering models
and their search is often a challenging task
\cite{LeonovKKSZ-2014,LeonovKYY-2015-TCAS,Kuznetsov-2016,DudkowskiJKKLP-2016,KuznetsovLYY-2017-CNSNS,LeonovKKM-2017}.
}
about the existence of periodic orbits.
In his paper \cite{Keldysh-1944} Keldysh wrote: ``{\it we do not give a rigorous mathematical proof ..., we construct a number of conclusions on intuitive considerations ...}''.

Nowadays we can apply rigorous analytical and reliable numerical methods,
which have been developed from 1944 till now:
theory of differential inclusions (see, e.g. \cite{Filippov-1988,GeligLY-1978});
direct Lyapunov method and frequency methods
(see, e.g. \cite{LurjeP-1944,GeligLY-1978,LeonovPS-1996});
numerical algorithms for solving differential inclusions
(see, e.g. \cite{AizermanP-1974,DontchevL-1992,PiiroinenK-2008}).
Follow the theory of differential inclusion,
for the model (\ref{sysK2d})
we consider the discontinuity manifold: $S=\{\dot x\!:\ \dot x=0\}$
on the phase space $(x, \dot x)$,
define $\varphi(\dot x)$ on $S$ as the set $[-\Phi,+\Phi]$,
and get differential inclusion
\begin{equation}\label{sysK2d-DI}
  J\ddot x+kx+\mu\dot x \in -\hat\varphi(\dot x),
  \quad
  \hat\varphi(\dot x) =
  \left\{
  \begin{array}{lc}
    \varphi(\dot x) & \dot x \neq 0, \cr
    [-\Phi,+\Phi] & \dot x = 0.
  \end{array}
  \right.
\end{equation}
The solutions of (\ref{sysK2d-DI}) are considered in the sense of Filippov \cite{Filippov-1988}.
Remark that here solutions cannot slide on the discontinuity manifold $S$,
but can tend to the stationary segment:
\[
  \Lambda=\{\ -\Phi/k \leq x \leq \Phi/k, \dot x=0\} \subset S,
\]
or pierce the manifold $S\backslash\Lambda$.

{\bf Theorem 1}. {\it If
\begin{equation}\label{2dstab-we}
   -2\sqrt{\Phi\kappa}<\mu, 
\end{equation}
then any solution of (\ref{sysK2d-DI}) converges to the stationary segment $\Lambda$.
}

{\it Proof sketch.}
Consider Lyapunov function
\[
  V(x,\dot x) = \frac{1}{2}(J\dot x^2 + k x^2).
\]
Since $\dot x \varphi(\dot x) >0$ for $\dot x \neq 0$,
we have
\[
  \dot V(x,\dot x) = - \mu\dot x^2 -\dot x\varphi(\dot x) <0,
  \quad \forall \dot x \notin S.
\]
The equality $V(x(t),\dot x(t)) \equiv {\rm const}$ can hold only for $x \in \Lambda$.
By the analog direct Lyapunov method for differential inclusions
\cite[Lemma 1.5, p.58]{GeligLY-1978}
we get the assertion of the theorem. $\blacksquare$

Thus,
rigorous analytical estimate (\ref{2dstab-we}) is close to
Keldysh's estimate of the global stability region (\ref{2dKstab}):
\(
  \delta_0=2\sqrt{\Phi\kappa}< 2.8\sqrt{\Phi\kappa}=\delta_K.
\)

Next, we consider Keldysh's estimate of the bounded stability region (\ref{2dK2lc})
and demonstrate the difficulties of the DFM application
to the systems with dry friction and stationary segment.
To simplify numerical simulation\footnote{
For numerical integration of differential inclusions we use \cite{PiiroinenK-2008}.
}
of (\ref{sysK2d-DI}) we use the following transformations:
\(
  x \to \frac{J}{\kappa} x, t \to  t \frac{\sqrt{J}}{\sqrt{k}} \ \Rightarrow \ J=1, k=1, \kappa=1.
\)

\begin{figure}[h]
 \centering
 \includegraphics[width=0.35\textwidth]{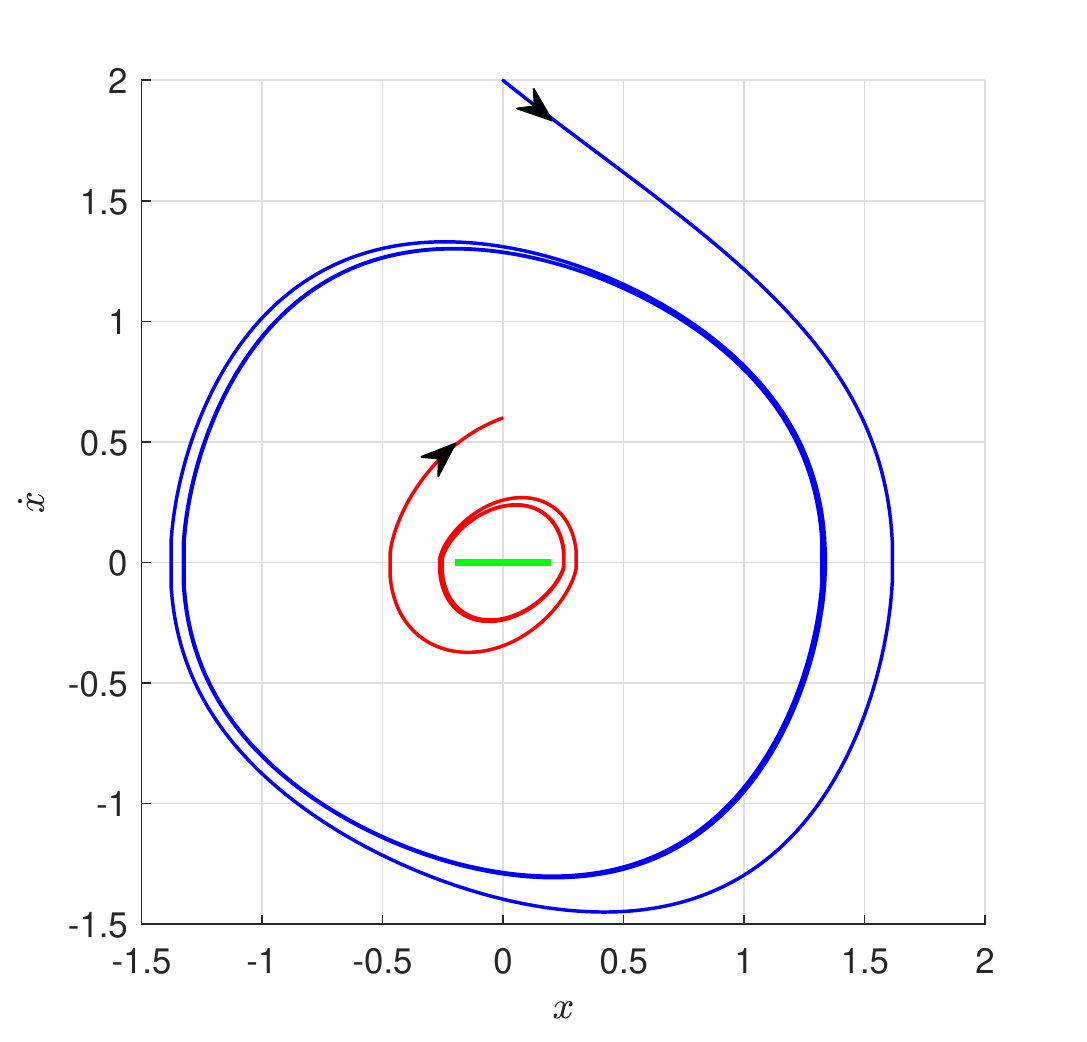}
 \includegraphics[width=0.35\textwidth]{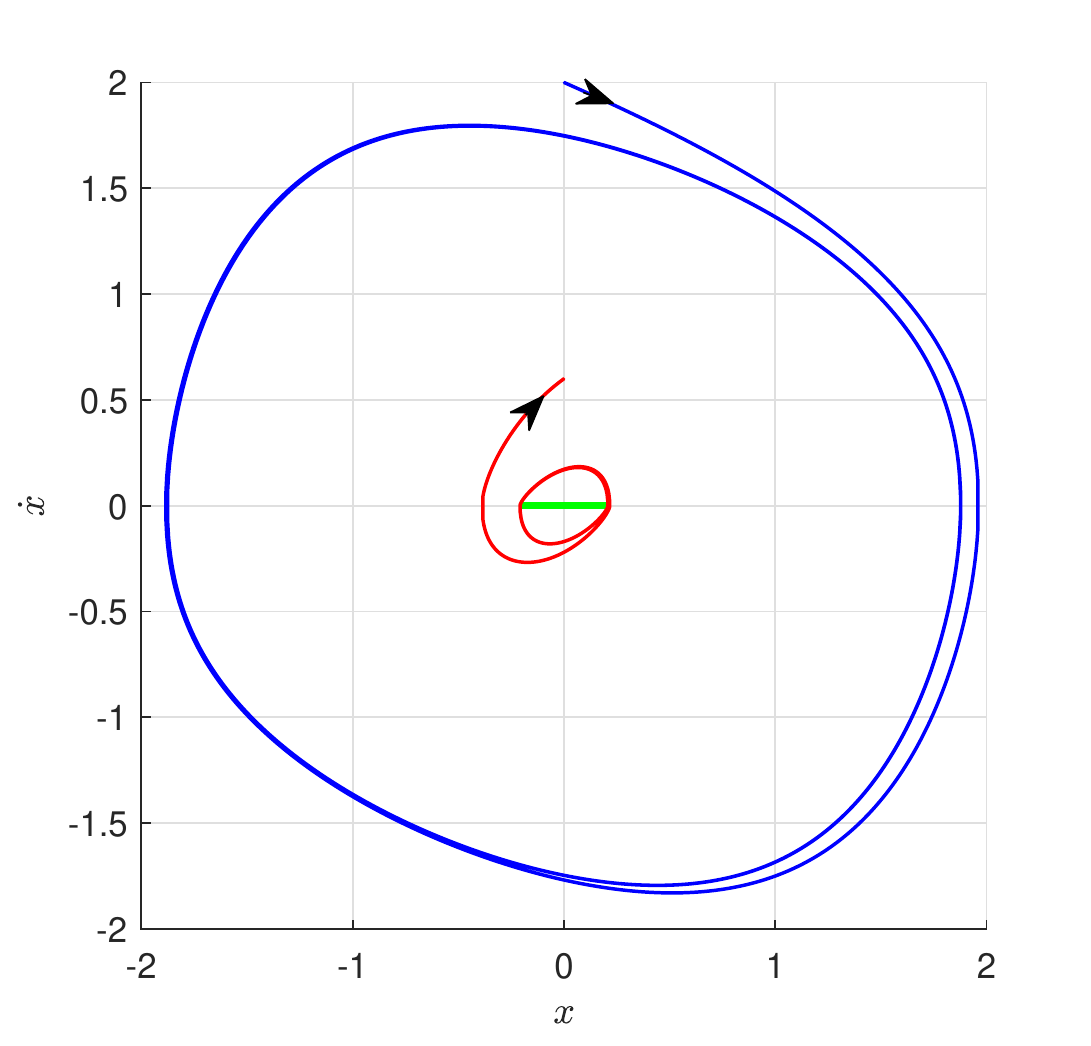}
 \caption{
 Numerical experiment with $\Phi=0.2$.
 Outer trajectory winds onto stable limit cycle,
 inner trajectory unwinds from unstable limit cycle
 and winds onto the stable limit cycle (hidden attractor).
 Left subfigure: $\mu\!=\!-1.3967\delta_{\!K}\!\!:a_+(\mu)\!>\!\!>\!a_-(\mu)\!>\!\Phi$.
 Right subfigure: $\mu\!=\!-1.7847\delta_{\!K}\!\!:a_+(\mu)\!>\!\!>\!\Phi\!>\!a_-(\mu)$.
 }\label{2d-phase}
\end{figure}

\noindent Fig.~\ref{2d-phase} shows qualitative behavior of trajectories
on the phase space in the case of multistability and coexistence of two limit cycles
for $\mu<-\delta_K$.
Here the external limit cycle is a hidden attractor
and corresponds to the flutter.
The basin of attraction of the stationary segment $[-\Phi,\Phi]$
is bounded by the internal unstable limit cycle.
When the harmonic approximation  $a_-\cos(t)$ of the unstable limit cycle
has the amplitude $a_- <\Phi$,
the stiffness of the model in vicinity of the ends of the stationary segment
complicates the modeling (the right subfigure of Fig.~\ref{2d-phase}).

\begin{figure}[h]
 \centering
 \includegraphics[width=0.35\textwidth]{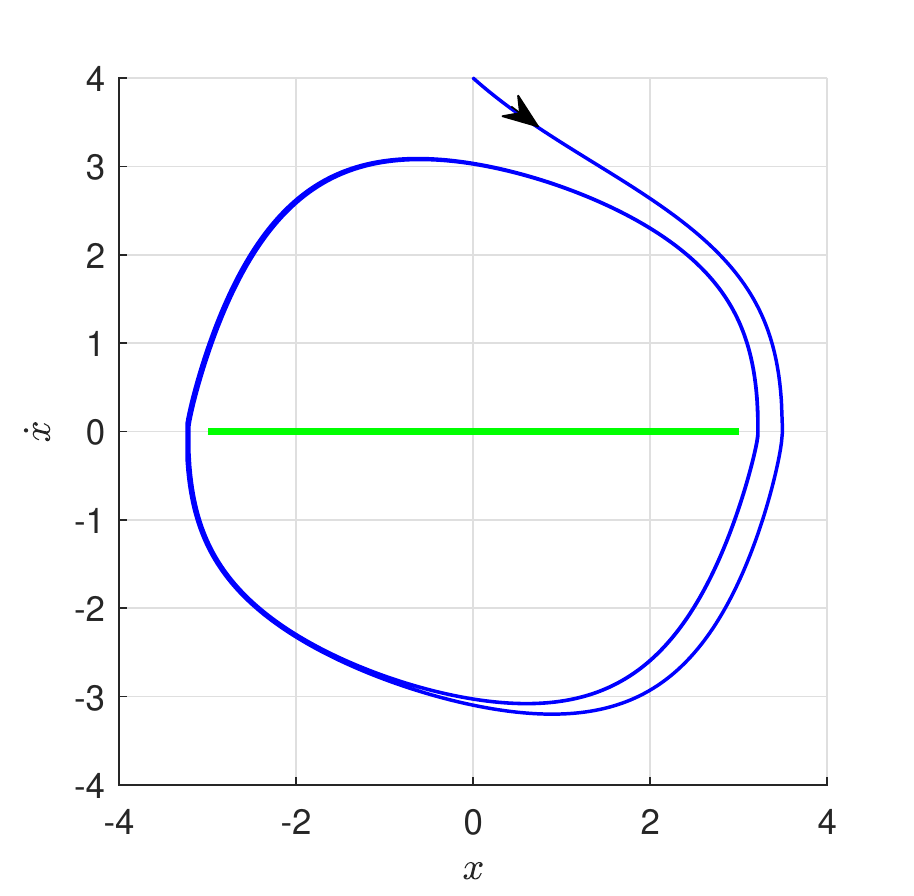}
 \includegraphics[width=0.35\textwidth]{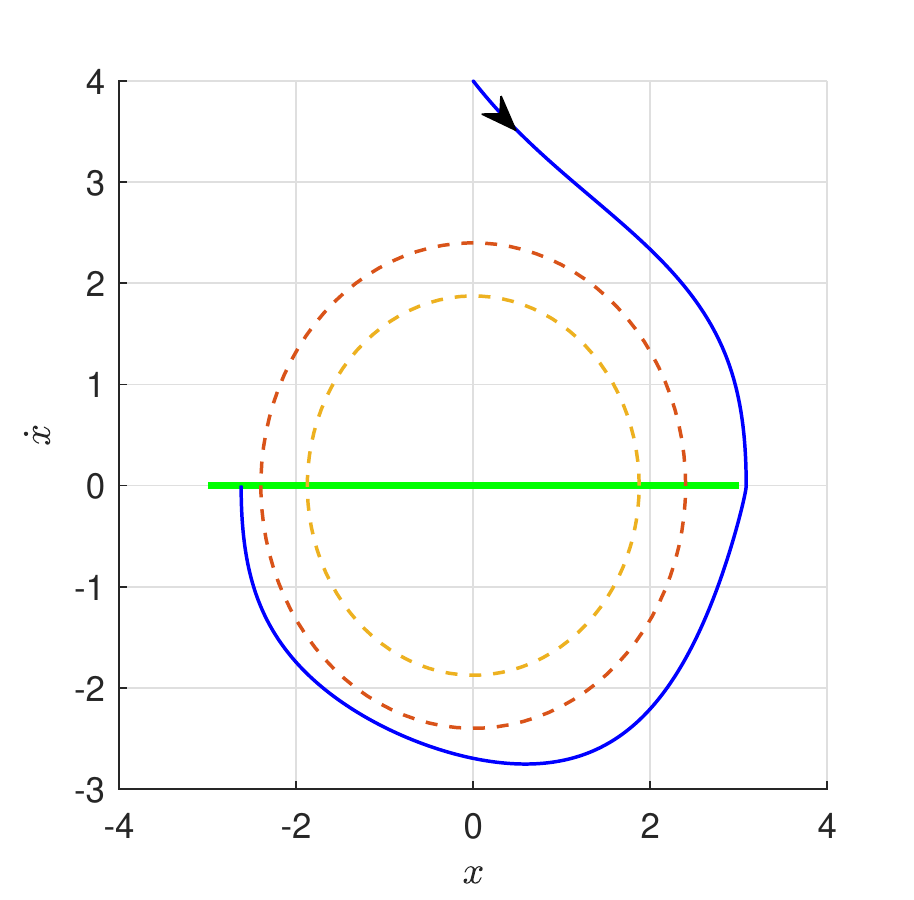}
 \caption{
 Numerical experiment with $\Phi\!=\!3$.
 Left subfigure: $\mu = -1.0713\delta_K$, $a_+(\mu) \gtrapprox \Phi> a_-(\mu)$;
 outer trajectory winds onto stable limit cycle,
 internal unstable limit cycle is not revealed numerically (due to stiffness).
 Right subfigure: $\mu = -1.0076\delta_K$,
 $\Phi \gtrapprox a_+(\mu)>a_-(\mu)$ (dash circles);
 outer trajectory approaches the stationary segment,
 both limit cycles have disappeared.
 }\label{2d-phase-1c}
\end{figure}

\noindent Fig.~\ref{2d-phase-1c} shows the bifurcation of collision of the external limit cycle and
the stationary segment.
In this numerical experiment both limit cycles have disappeared,
while Keldysh's estimate (\ref{2dK2lc}) holds.

Thus, Keldysh's estimate of the stability domain
may be far from the numerical estimates.

\section{CONCLUSIONS}
Similar model with two degrees of freedom,
considered by Keldysh in his paper \cite{Keldysh-1944},
can be rigorously studied in the same way.
For this case, the existence of the Lyapunov function
and stability in large
can be effectively proved
by special frequency criteria \cite{LeonovK-2018-DAN}.
Nowadays the study of stability in large and oscillations
for modern aircrafts is also motivated by such problems
as pilot induced oscillations
and actuator saturations 
(see, e.g. surveys \cite{LeonovAKP-2012,AndrievskyKL-2017} and references within).
A terrifying illustration of such effects is the YF-22
crash in April 1992 and Gripen crash in August 1993 \cite{Dornheim92,Shifrin93}.
In these regards, in the work \cite{LauvdalMF-1997} it is written that
``{\it since stability in simulations does not imply
stability of the physical control system,
stronger theoretical understanding is required}''.

\section{ACKNOWLEDGMENTS}
Authors would like to thank Olga F. Radzivill (TsAGI Library),
Boris R. Andrievsky (IPME RAS).
This work was supported by the grant NSh-2858.2018.1 
for the Leading Scientific Schools of Russia (2018-2019).

\bibliographystyle{aipnum-cp}

%

\end{document}